\documentclass[sigconf]{acmart}

\AtBeginDocument{%
  }

\setcopyright{acmcopyright}
\copyrightyear{2024}
\acmYear{2024}
\acmDOI{XXXXXXX.XXXXXXX}

\acmConference[Conference acronym 'XX]{Make sure to enter the correct
  conference title from your rights confirmation emai}{June 03--05,
  2024}{Woodstock, NY}
\acmPrice{15.00}
\acmISBN{978-1-4503-XXXX-X/18/06}

\usepackage{bm} % 加粗支持
\usepackage{amsmath,amsfonts}
\usepackage{algorithmic}
\usepackage{graphicx, subfigure, verbatim, url,pifont}
\usepackage{stmaryrd}
\usepackage[ruled,vlined]{algorithm2e}

 \usepackage{multirow}
 \usepackage{booktabs}
\usepackage{makecell}

\usepackage{subfloat}

\begin{document}

\title{CDIO: Cross-Domain Inference Optimization with Resource Preference Prediction for Edge-Cloud Collaboration}

% \author{Submission ID: 226}
% \affiliation{%
%   \institution{Anonymous Authors}
%   \institution{Anonymous Affiliations}
%   % \institution{University of Chinese Academy of Sciences}
%  % \streetaddress{1 Th{\o}rv{\"a}ld Circle}
%  % \city{Beijing}
%   \country{Anonymous Countries} \\
%  \textcolor{white}{  \institution{Anonymous Authors}
%   \institution{Anonymous Affiliations}
%   % \institution{University of Chinese Academy of Sciences}
%  % \streetaddress{1 Th{\o}rv{\"a}ld Circle}
%  % \city{Beijing}
%   \country{Anonymous Countrys}\\
%    \country{Anonymous Countrys}\\
%    \country{Anonymous Countrys}}
%   }

\author{Zheming Yang\textsuperscript{\rm 1,2,4},
    Wen Ji\textsuperscript{\rm 1,3},
    Qi Guo\textsuperscript{\rm 1,2},
    Dieli Hu\textsuperscript{\rm 1,2,3},
    Chang Zhao\textsuperscript{\rm 1,2},
    Xiaowei Li\textsuperscript{\rm 1},
     Xuanlei Zhao\textsuperscript{\rm 4},
    Yi Zhao\textsuperscript{\rm 4,5},
     Chaoyu Gong\textsuperscript{\rm 4} and
    Yang You\textsuperscript{\rm 4}\\}
\affiliation{%
  \textsuperscript{\rm 1}\institution{Institute of Computing Technology, Chinese Academy of Sciences}
\textsuperscript{\rm 2}\institution{University of Chinese Academy of Sciences}
\textsuperscript{\rm 3}\institution{Peng Cheng Laboratory}
% \textsuperscript{\rm 4} \institution{Lishui Institute of Hangzhou Dianzi University}
 \textsuperscript{\rm 4}\institution{National University of Singapore}
 \textsuperscript{\rm 5}\institution{Tsinghua University}
 \textcolor{white}{ \country{China}}
  }

\renewcommand{\shortauthors}{Anonymous Authors.}

\begin{abstract}
Currently, massive video tasks are processed by edge-cloud collaboration. However, the diversity of task requirements and the dynamics of resources pose great challenges to efficient inference, resulting in many wasted resources. In this paper, we present CDIO, a cross-domain inference optimization framework designed for edge-cloud collaboration. For diverse input tasks, CDIO can predict resource preference types by analyzing spatial complexity and processing requirements of the task. Subsequently, a cross-domain collaborative optimization algorithm is employed to guide resource allocation in the edge-cloud system. By ensuring that each task is matched with the ideal servers, the edge-cloud system can achieve higher efficiency inference.  The evaluation results on public datasets demonstrate that CDIO can effectively meet the accuracy and delay requirements for task processing. Compared to state-of-the-art edge-cloud solutions, CDIO achieves a computing and bandwidth consumption reduction of 20\%-40\%. And it can reduce energy consumption by more than 40\%.
\end{abstract}

\begin{CCSXML}
<ccs2012>
 <concept>
  <concept_id>10010520.10010553.10010562</concept_id>
  <concept_desc>Computer systems organization~Embedded systems</concept_desc>
  <concept_significance>500</concept_significance>
 </concept>
 <concept>
  <concept_id>10010520.10010575.10010755</concept_id>
  <concept_desc>Computer systems organization~Redundancy</concept_desc>
  <concept_significance>300</concept_significance>
 </concept>
 <concept>
  <concept_id>10010520.10010553.10010554</concept_id>
  <concept_desc>Computer systems organization~Robotics</concept_desc>
  <concept_significance>100</concept_significance>
 </concept>
 <concept>
  <concept_id>10003033.10003083.10003095</concept_id>
  <concept_desc>Networks~Network reliability</concept_desc>
  <concept_significance>100</concept_significance>
 </concept>
</ccs2012>
\end{CCSXML}

\ccsdesc[500]{Information systems~Multimedia streaming}
\ccsdesc[300]{Computer systems organization}

\keywords{Multimedia system design; edge-cloud collaborative inference; video processing; resource optimization}


\maketitle

\section{Introduction}
With the rapid development of deep learning and Internet of Things (IoT) technologies, the scale of connected IoT devices continues to expand \cite{li2018learning404040}. Various types of end devices are gradually being applied in various scenarios \cite{ji2020crowd414141,yang2023visual111}, including smart cities, intelligent transportation, and industrial IoT. The massive volume of image and video data generated by a large number of end devices requires real-time transmission and analysis. Many DNN models are deployed on servers for real-time inference \cite{du2020server222}. Although cloud servers have significant computing resources, processing all inference tasks in the cloud results in large-scale data transmission, which limits the performance of real-time inference due to limited bandwidth resources \cite{hua2023edge333}. In addition, edge devices with limited computing resources can only deploy lightweight DNN models \cite{wang2023shoggoth444,zhou2019edge424242}, which can only support simple tasks. Accuracy may be compromised when handling complex tasks.

\begin{figure}[t!]
	\centering 
		\includegraphics[width=1\linewidth]{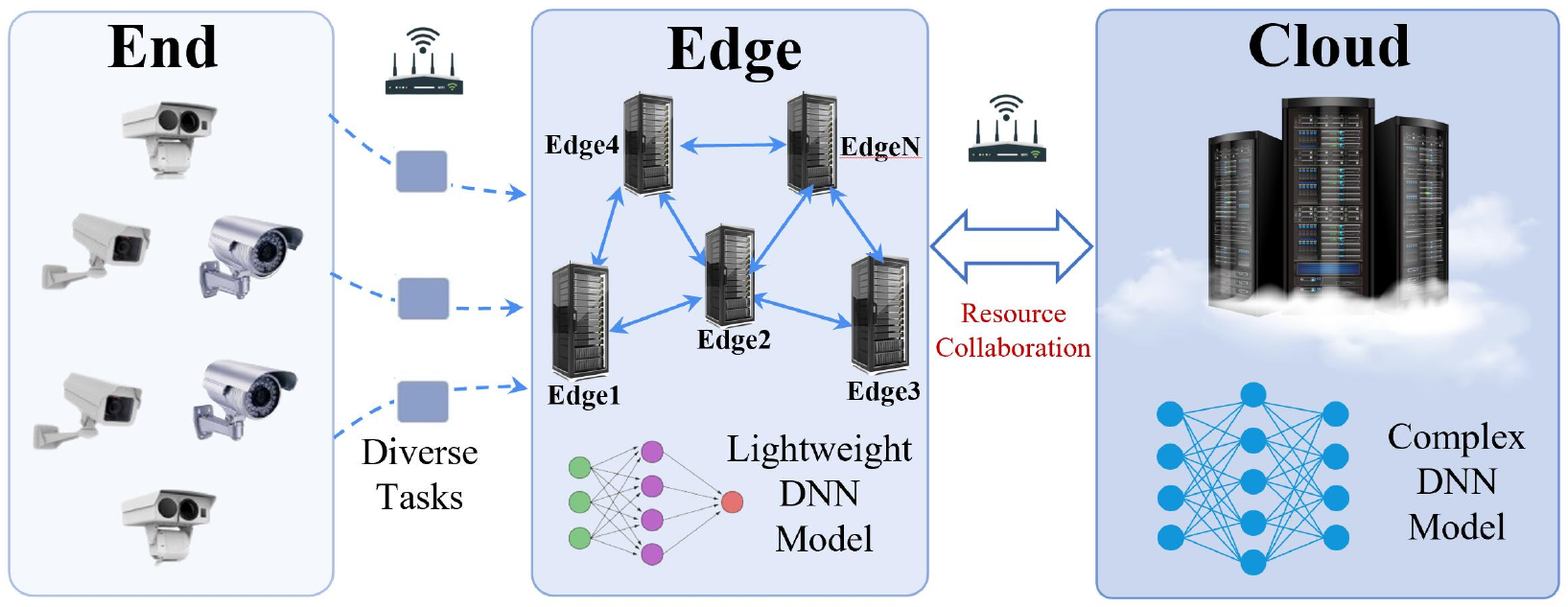}
	\caption{The illustration of edge-cloud collaborative inference architecture.}
	\label{figure1}
\end{figure}

\par
\begin{figure}[t]
	\centering 
    \subfigure[Edge, \textit{Task 1}]{
    \label{Fig.sub.2.1}
    \includegraphics[width=0.23\textwidth]{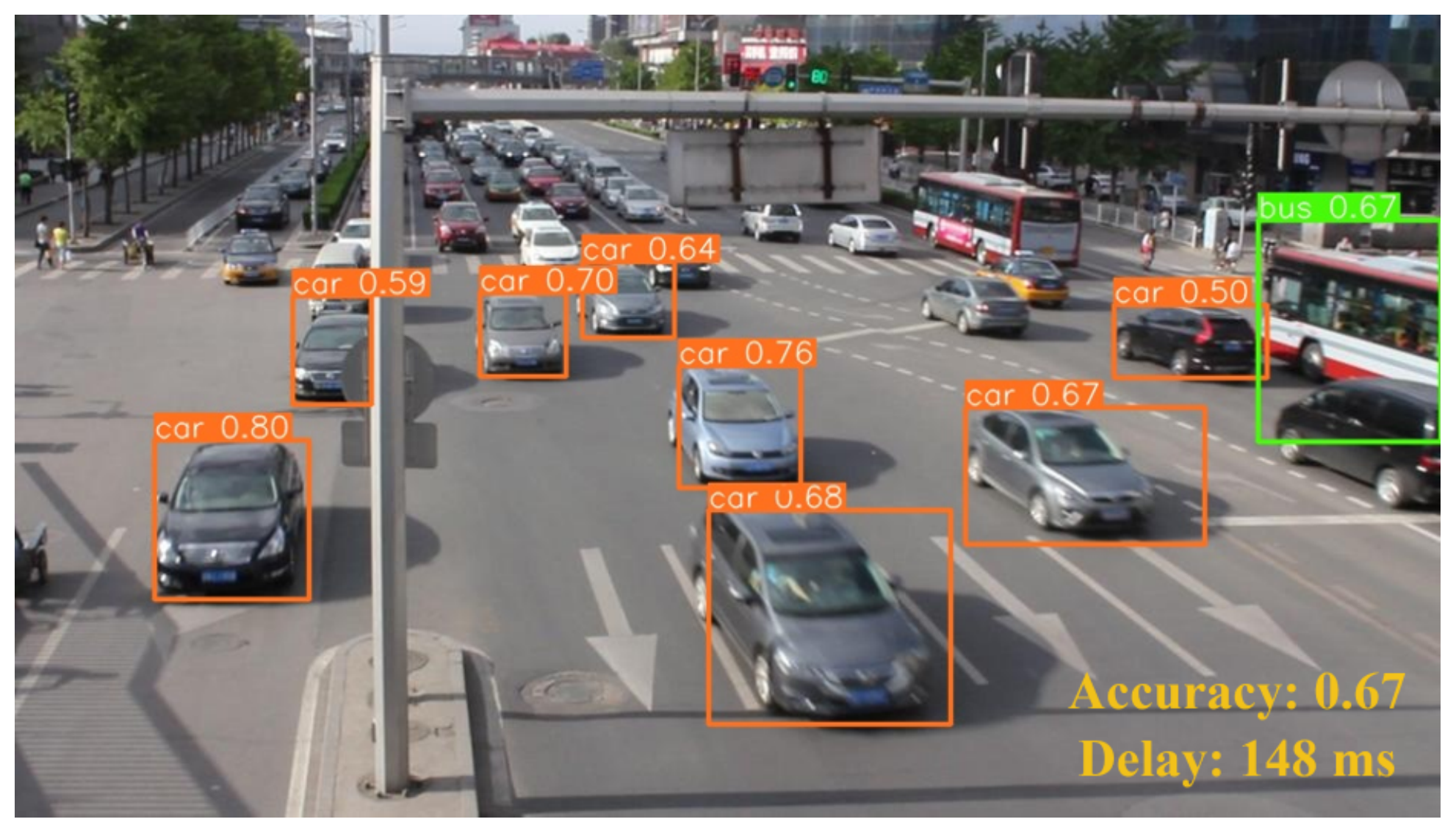}}
    \subfigure[Cloud, \textit{Task 1}]{
    \label{Fig.sub.2.2}
    \includegraphics[width=0.23\textwidth]{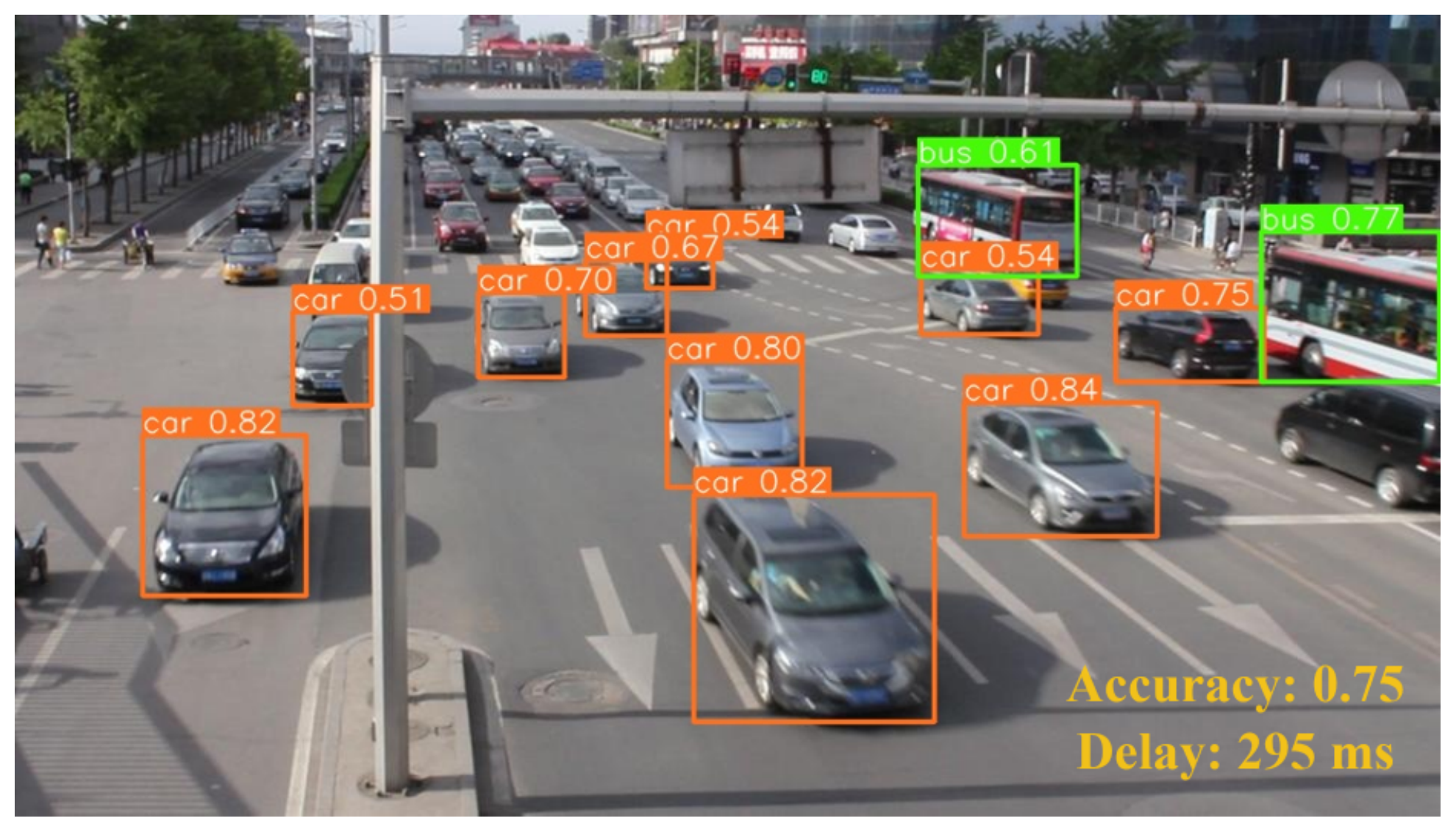}}
        \subfigure[Edge, \textit{Task 2}]{
    \label{Fig.sub.2.3}
    \includegraphics[width=0.23\textwidth]{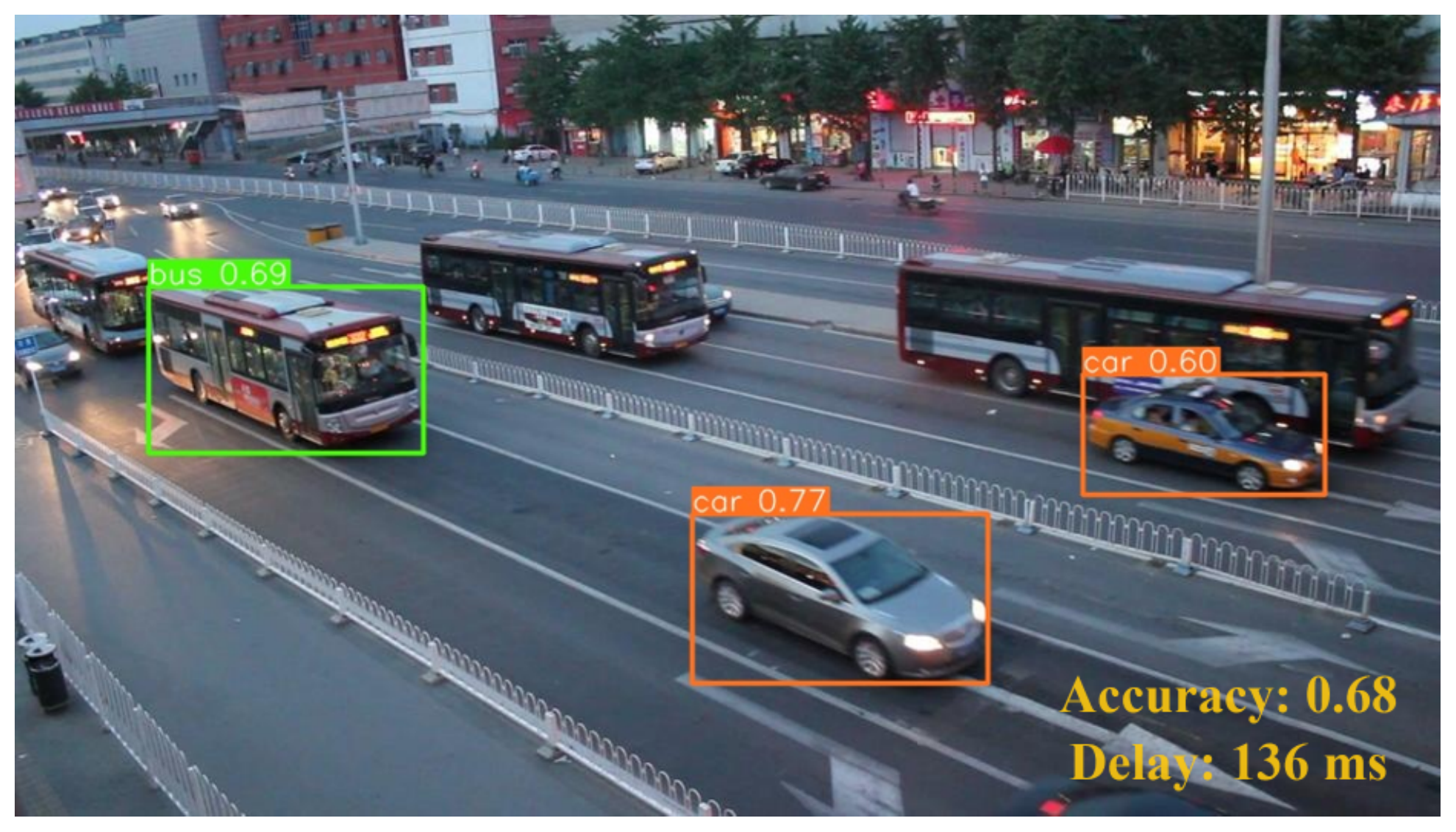}}
            \subfigure[Cloud, \textit{Task 2}]{
    \label{Fig.sub.2.4}
    \includegraphics[width=0.23\textwidth]{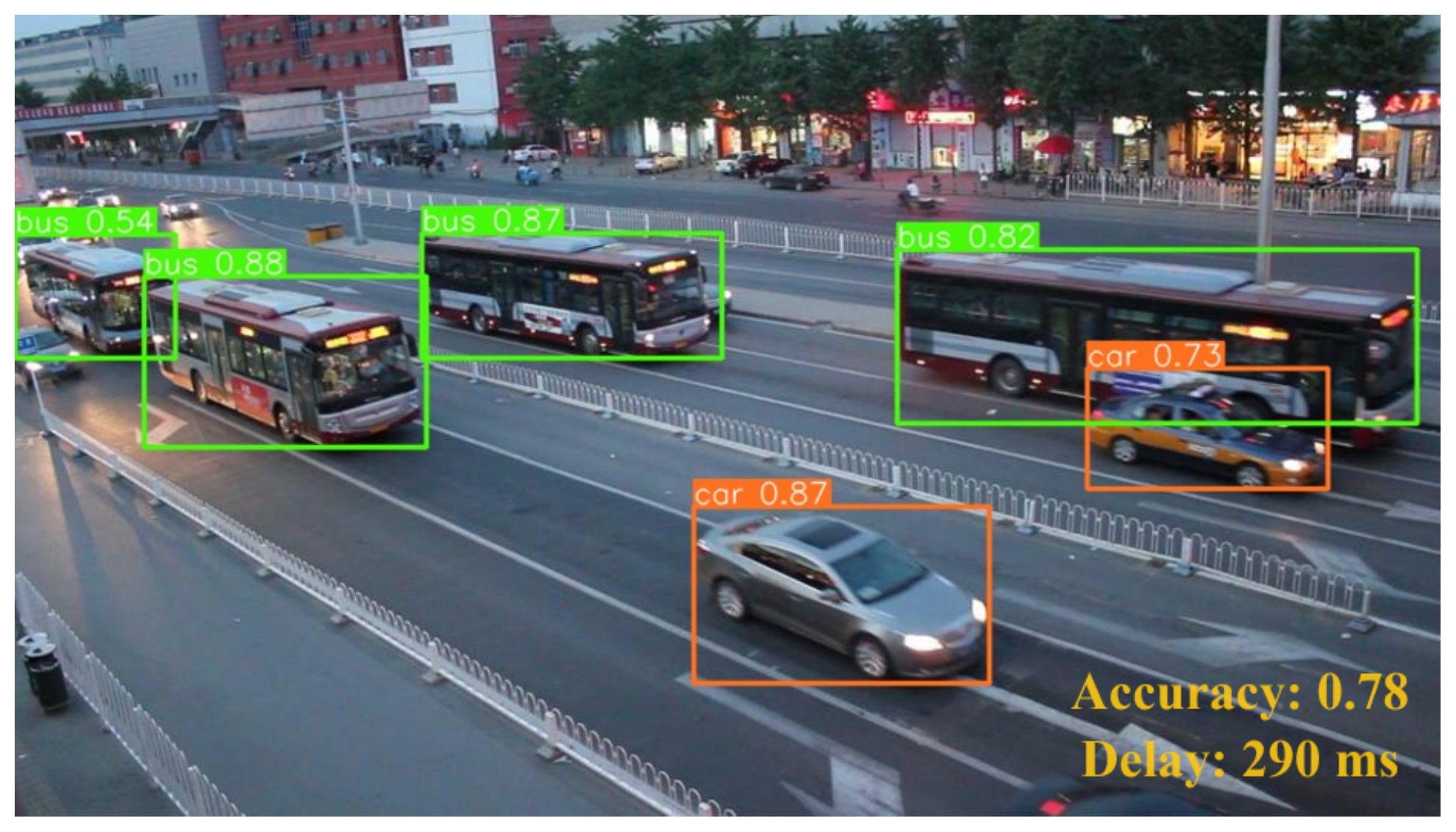}}
    
    \caption{The inference results for different tasks on edge and cloud. If \textit{Task 1} is more sensitive to delay, it is suitable for uploading to the edge processing. If \textit{Task 2} requires better accuracy, it is suitable for processing in the cloud.}
	\label{figure2}
\end{figure}

Recently, to achieve more efficient inference, many researchers explore collaborative inference architectures \cite{duan2022distributed555,li2021appealnet454545,yang2022cnnpc464646} based on edge-cloud systems, as illustrated in Figure~\ref{figure1}. This architecture aims to leverage the strengths of both edge and cloud computing. However, despite its potential benefits, this architecture also presents several challenges that need to be addressed. One of the key challenges lies in the diverse requirements that different tasks impose on various types of resources within the system. For instance, some tasks are highly sensitive to delay and thus require more bandwidth resources to ensure real-time data transmission and processing \cite{meng2019dedas434343}. Some tasks prioritize accuracy and therefore require additional computational resources to execute complex DNN models. Figure~\ref{figure2} shows an example of the processing of two tasks. Assume that \textit{Task 1} is more sensitive to delay and \textit{Task 2} requires better accuracy. We find that different resource allocation strategies have a huge impact on the inference process of edge-cloud collaboration. It is obvious that \textit{Task 1} is suitable for uploading to the edge processing and \textit{Task 2} is suitable for processing in the cloud with sufficient computational resources. This inherent diversity in task requirements increases the difficulty of effective collaborative inference strategies within edge-cloud systems \cite{gazzaz2022croesus202020}. Moreover, in an edge-cloud collaborative system, the dynamism and uncertainty of resources make it more difficult to allocate tasks rationally \cite{wang2018dynamic777}. Resource availability can fluctuate unpredictably due to factors such as network congestion and varying workloads \cite{kherraf2019optimized444444}. This poses significant challenges for collaborative inference under dynamic resource conditions in edge-cloud systems.

\par
To address the above problems, we propose a cross-domain inference optimization framework for edge-cloud collaboration, named CDIO. For real-time video analysis tasks, CDIO can analyze the type of resource preference through the pre-processor. It then optimizes the resource allocation of the edge-cloud system to meet the task processing requirements while minimizing resource consumption. The main goal of CDIO is to achieve adaptive scheduling of diverse inference tasks under dynamic resource conditions and enhance the efficiency of video inference. The main contributions of this paper can be summarized as follows.

\begin{itemize}
\item A resource preference prediction model based on spatiotemporal feature analysis is proposed. It can predict resource preference type by analyzing the complexity and requirements of tasks. It provides guiding information for resource allocation in edge-cloud systems.

\item A cross-domain collaborative optimization algorithm based on a feedback mechanism is developed. Combining the resource preference types of tasks,  the algorithm can optimize the resource allocation in the edge-cloud system under dynamic resource conditions.

\item The performance of the proposed framework is evaluated and compared with the baseline method. The experimental results show that CDIO can reduce the computing and bandwidth consumption of video processing by 20\%-40\% and reduce energy consumption by more than 40\%.
\end{itemize}

\par
The remainder of this paper is organized as follows. We first summarize the related work in Section 2. In Section 3, the proposed CDIO framework is described in detail. We evaluate our proposed method through extensive experiments in Section 4. Section 5 discusses the solutions in this paper. Section 6 concludes the paper and presents future work.

\section{Related Work}
\textbf{Edge-Cloud Collaboration for Resource Prediction.} 
The main goal of resource prediction is to optimize resources in the long term \cite{sun2016optimizing353535}. Through resource prediction, edge-cloud systems can allocate resources well in advance to address potential high loads or performance bottlenecks in the future \cite{miao2020intelligent363636}. Some researchers propose edge-cloud collaboration solutions based on resource prediction. The authors in \cite{kang2017neurosurgeon222222,dong2022splitnets232323} focus on model segmentation schemes based on edge-cloud collaboration and determining the optimal model segmentation point according to computing resources. However, they are only for single-edge nodes. The authors in \cite{yang2023javp666} propose a joint-aware video processing architecture for edge-cloud collaboration. It can guide resource allocation and reduce video processing costs by predicting the task complexity. Considering the constraints of multiple tasks and edge resources, the authors in \cite{sun2021cloud121212} utilize gated recurrent units to predict edge resource utilization. They then propose a joint optimization method based on resource utilization prediction according to the predicted results of network states. To address the load imbalance problem in edge-cloud systems, the authors in \cite{li2023elastic131313} propose a resource optimization method based on workload prediction, which improves prediction accuracy through server correlation analysis. The authors in \cite{fang2018nestdnn141414} introduce a dynamic resource prediction framework for DNN models, allowing the selection of the optimal balance between resources and accuracy for each DNN model. The work in \cite{gujarati2020serving242424} adjusts resource allocation schemes by predicting model inference times. The authors in \cite{kim2023moca252525} dynamically adjust memory access rates based on delay targets and user-defined priorities to improve resource allocation efficiency. However, the above works only consider the prediction and optimization of a single type of resource (only computing or only bandwidth). The characteristic differences in heterogeneous resources may cause uneven resource utilization \cite{feng2022heterogeneous373737}, thus affecting the performance of edge-cloud systems.

\par
\textbf{Edge-Cloud Collaboration for Inference Offloading.} The offloading of video tasks from the end side to the edge or cloud servers is a crucial step for efficient video processing \cite{grulich2018collaborative383838}. Some researchers propose many edge-cloud collaboration solutions based on task offloading. The authors in \cite{liu2022sniper888} introduce a time-aware edge-cloud collaborative task scheduling method, which ensures scheduling accuracy and improves throughput by the performance characterizing network. The authors in \cite{yangadaptive262626} propose an edge-cloud collaborative scheduling framework for joint configuration optimization. It can improve task allocation through two-stage robust optimization. The authors in \cite{nigade2022jellyfish272727} dynamically adjust the task offloading scheme through service prioritization and network conditions. The authors in \cite{murad2022dao999} propose a dynamic adaptive offloading framework for video analysis, which enhances inference accuracy by dynamically adjusting network bandwidth and video bitrate. Simultaneous uploading of many tasks can affect the accuracy and real-time performance of video processing. To address this problem, the authors in \cite{xu2020energy101010} investigate the dynamic task offloading problem for large-scale inference requests and design an online optimization algorithm that supports real-time adjustments. Considering the uncertainty in task arrival rates, the authors in \cite{111111} propose a dual time-scale Lyapunov optimization algorithm to overcome the uncertainty of future information of the system, aiming to minimize the cost of task offloading. To solve the load imbalance problem with multiple edge nodes, the authors in \cite{tang2020deep282828,qu2021dmro292929} investigate task offloading schemes based on deep reinforcement learning. In real-world scenarios, due to the continuous change of accuracy and delay requirements of tasks, different tasks have different preferences for different types of resources \cite{kumar2015preference393939}. The above methods ignore analyzing task characteristics, which makes it difficult to achieve efficient edge-cloud collaborative task offloading.

\section{The Proposed CDIO Framework}
In this section, we present CDIO, a cross-domain inference optimization framework for edge-cloud collaboration. Section 3.1 shows the overall design of CDIO. Then the resource preference prediction module is introduced in Section 3.2, and the cross-domain collaborative optimization module is described in Section 3.3.

\subsection{Overall Design}
\begin{figure}[t!]
	\centering 
		\includegraphics[width=1\linewidth]{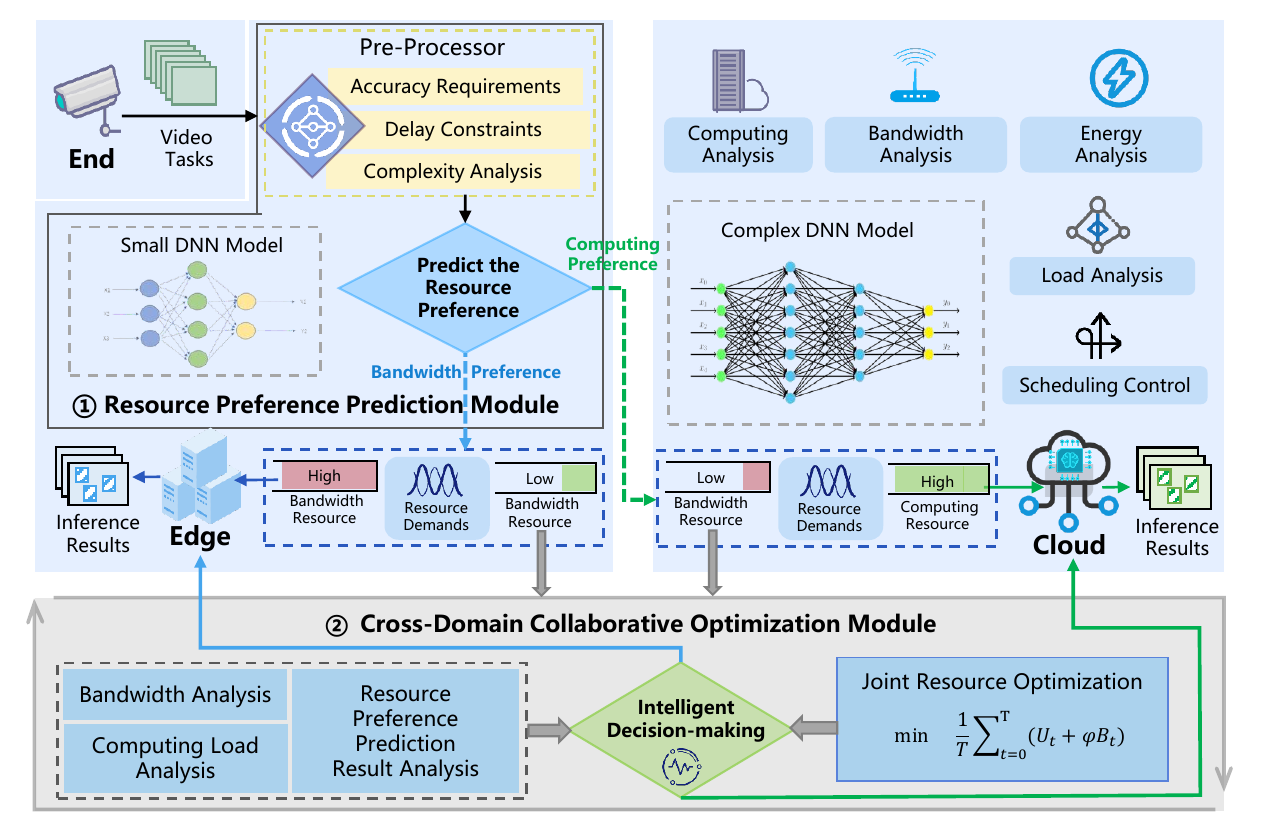}
	\caption{The workflow of the proposed CDIO framework.}
	\label{figure3}
\end{figure}

Figure~\ref{figure3} illustrates the overall design and workflow of the CDIO framework,  including the end, edge, and cloud components. Before using the framework, a lightweight DNN model needs to be deployed on the edge, and a complex DNN model needs to be deployed in the cloud for real-time inference of end input tasks \cite{ali2020res343434}. During the application process, tasks sequentially pass through two modules of CDIO, namely the resource preference prediction module in Section 3.2 and the cross-domain collaborative optimization module in Section 3.3. The resource preference prediction module first analyzes factors such as the complexity, accuracy requirements, and delay constraints of the input task. Subsequently, it uses a Long Short-Term Memory (LSTM) model based on spatiotemporal feature analysis to predict the resource preference type for the task. If the task prefers computing resources, it is prioritized for inference on cloud servers. Otherwise, if the task prefers bandwidth resources, it is prioritized for inference on edge devices. However, it's important to note that these allocations are not rigidly enforced and are subject to adjustments based on the current resource conditions within the system. The cross-domain collaborative optimization module dynamically determines the resource allocation strategy based on the results of resource preference prediction and the current computing load and bandwidth conditions of the system. Overall, this framework aims to achieve high-performance edge-cloud collaborative inference with as few resources as possible. The adaptability of the system to changes in task requirements and resource availability is a critical aspect, ensuring that the framework remains efficient in varying scenarios.

\subsection{Resource Preference Prediction}
The prediction of resource preferences for input tasks can provide decision-making information for edge-cloud collaboration \cite{pang2020adaptive333333}. In this context, we propose a resource preference prediction model based on spatiotemporal feature analysis. This model determines resource preference types by analyzing features such as spatial complexity, accuracy requirements, and delay requirements of tasks. First, we organize historical data in a time series to ensure that each time point includes features of previous tasks and their corresponding resource preference types. Renormalization group method \cite{bagrov2020multiscale151515} has been proven to quantitatively describe the spatial complexity of images by associating information across different scales. A new feature binary is formed by introducing the location information of image pixels, denoted as $(a, e)$. Where $a$ denotes the row index of the pixel, $e$ denotes the column index of the pixel. The degree of overlap between images at different scales is calculated as follows:
\begin{equation}
\begin{aligned}
O_{n, n-1} =\frac{G^2}{L_{n-1}^2} \sum_{a=1}^{L_n} \sum_{e=1}^{L_n} \mathbf{s}_{a e}^2(n)=\frac{G^2}{L_{n-1}^2} \cdot L_n^2 \cdot O_{n, n}
\end{aligned}
\end{equation}

The initial image is denoted as $n=0$, and $O_{n, n-1}$ signifies the overlap between the image at scale $n$ and scale $n-1$. In each iteration of the renormalization process, the image undergoes partitioning into blocks with dimensions $G \times G$. Subsequently, each of these blocks is substituted with a single pixel whose state is determined by $\mathbf{s}_{a e}(n)$, where $L_n$ indices enumerate the pixels within the same block. Considering the features appearing on each subsequent scale, we can obtain the space complexity of the image as follows:

\begin{equation}
C=\sum_{n=0}^{N-1} C_n=\sum_{n=0}^{N-1}\left|O_{n+1, n}-\frac{1}{2}\left(O_{n, n}+O_{n+1, n+1}\right)\right|
\end{equation}

Next, to better capture the spatiotemporal features and processing requirements of the task, we design the following model:

\begin{figure}[t!]
	\centering 
		\includegraphics[width=1\linewidth]{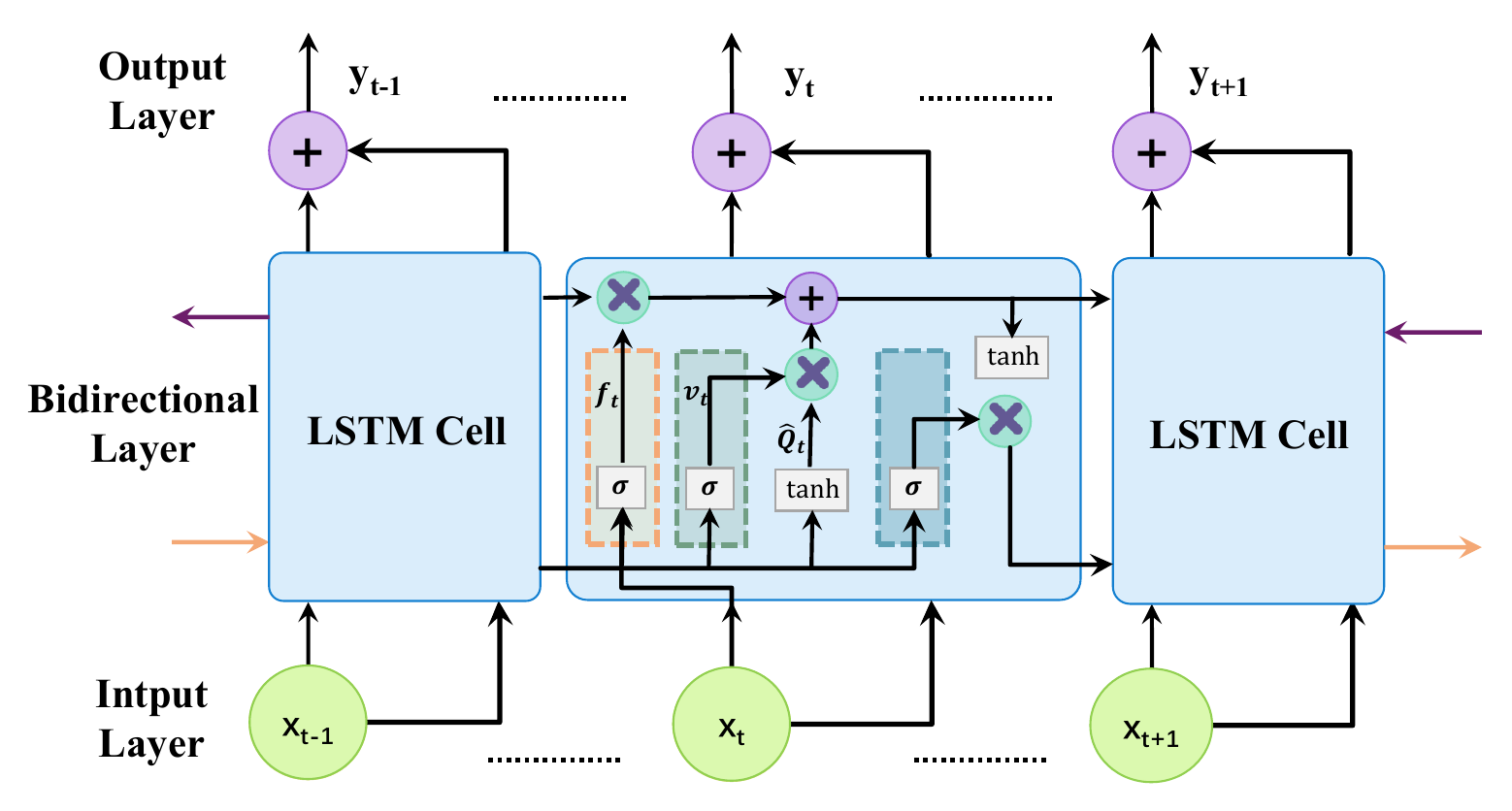}
	\caption{The overview of the multi-layer LSTM model.}
	\label{figure4}
\end{figure}

\begin{equation}
x_t=I_t+C_t+A_t+D_t
\end{equation}
where $I_t$ denotes the time linearity, while $C_t$ denotes the space complexity, $A_t$ is the accuracy requirement, and $D_t$ is the delay requirement. Then, we design a multi-layer LSTM model based on spatiotemporal feature analysis, as shown in Figure~\ref{figure4}. It combines spatiotemporal features with other features as inputs to the LSTM model where the length of the input sequence is determined by the historical information \cite{ouhame2021efficient191919}. We then use multi-layer LSTM units to enhance the learning capability of the model. Each layer can extract features of time series data at different levels \cite{su2020convolutional323232}. Each LSTM can process inputs for one time step and update its internal state. A basic LSTM network consists of a forget gate, an input gate, and an output gate. The formula for the forget gate is as follows:

\begin{equation}
f_t=\sigma_f\left(W_f \cdot\left[h_{t-1}, x_t\right]+b_f\right) 
\end{equation}

The $\sigma$ denotes the activation function, $W$ and $b$ denote the weight matrix and bias, respectively. $h_{t-1}$ denotes the output of the LSTM cell at the moment $t-1$, and $x_t$ represents the input at the moment $t$. The formula for the input gate is as follows:

\begin{equation}
\begin{gathered}
v_t=\sigma_i\left(W_v \cdot\left[h_{t-1}, x_t\right]+b_v\right) \\
\tilde{Q}=\tanh \left(W_Q \cdot\left[h_{t-1}, x_t\right]+b_Q\right)
\end{gathered}
\end{equation}
where $\tilde{Q}$ is the cell state. The cell state is updated by $Q_t=f_t \odot Q_{t-1}+v_t \odot \tilde{Q_t}$. $\tilde{Q_t}$ is the cell state at time step $t$ and $Q_{t-1}$ is the cell state at the previous time step. The formula for the output gate is as follows:

\begin{equation}
\begin{gathered}
o_t=\sigma_o\left(W_o \cdot\left[h_{t-1}, x_t\right]+b_o\right) \\
h_t=o_t \odot \tanh \left(Q_t\right)
\end{gathered}
\end{equation}
where $h_t$ is the hidden state of the current time step. The design of the output gate is conceptualized to address a binary classification problem, determining the predilection of the task for either computing or bandwidth resources.  To accomplish this, we use the Binary Cross-Entropy (BCE) as the loss function. The BCE loss function is formulated as follows:

\begin{equation}
B C E=-\frac{1}{K} \sum_{i=1}^K\left[g_i \log \left(\hat{g_i}\right)+\left(1-g_i\right) \log \left(1-\hat{g_i}\right)\right]
\end{equation}
where $g_i$ is the true value and $\hat{g_i}$ is the predicted value. The goal of the proposed LSTM model is to learn the complex relationship between task features and resource preferences. It can capture the temporal dependencies between task features to make effective predictions about the types of resource preferences for new tasks. This can provide a strong foundation for efficient resource allocation in edge-cloud systems.

\subsection{Cross-Domain Collaborative Optimization}
\subsubsection{Problem Formulation}
In edge-cloud systems, many video tasks need to be uploaded to the server for real-time inference \cite{khani2021real313131}. We denote the set of input tasks by $V=\left\{v_1, v_2, \cdots v_i, \cdots v_K\right\}$. And $Y=\left\{y_1, y_2, \cdots y_j, \cdots y_M\right\}$ is used to represent the set of servers, where $y_M$ denotes the cloud server. To optimize the allocation of computing and bandwidth resources and thereby enhance their overall utilization, we propose a comprehensive objective function. This function is a combined weighted sum that encapsulates both computing and bandwidth resources. The formulation of this objective function inherently incorporates the tradeoff between computing and bandwidth resources, ensuring a balanced consideration during the optimization process.  Subsequently, we impose accuracy and delay requirements as constraint conditions. The accuracy constraint ensures that the allocation mechanism does not sacrifice the quality of outcomes for resource efficiency, while the delay constraint guarantees that tasks are completed within an acceptable time.  This can be formalized as follows:

\begin{equation}
\begin{array}{ll}
 &  \min \quad  \frac{1}{T} \sum_{t=0}^{T}\left(U_{t}+ \varphi  B_{t}\right) \\

\text { s.t. } &  C_{1}: A_{i,t} \geq A_{i,t}^{q},i\in \left\{1, 2, \cdots K\right\}, t\in T \\

& C_{2}: D_{i,t} \leq D_{i,t}^{q},i\in \left\{1, 2, \cdots K\right\}, t\in T \\

& C_{3}: \sum_{j=1}^{M} z_{i, t}^{j}=1, z_{i, t}^{j}=\{0,1\},y_{j} \in Y, t\in T

\end{array}
\end{equation}

$U_{t}$ and $B_{t}$ are the computing resource consumption and bandwidth resource consumption for time slot $t$. $A_{i,t}$ is the accuracy of task $i$ and $A_{i,t}^{q}$ is the accuracy requirement. $D_{i,t}$ is the delay of task $i$ and $D_{i,t}^{q}$ is the delay requirement. $z_{i, t}^{j}$ indicates whether task $i$ is assigned to the $j$-th server. The weight parameter $\varphi$ is used to control the tradeoff between computing and bandwidth resources. Constraint $C_{1}$ and constraint $C_{2}$ are used to ensure that the accuracy and delay requirements of each task are met. Otherwise, it will be assigned to a more resource-rich server. Constraint $C_{3}$ ensures that only one server can be selected for each task at time slot $t$. Our goal is to minimize the resource consumption of the edge-cloud system while satisfying the task accuracy and delay requirements. This formulation not only aims at optimizing resource utilization but also imposes essential constraints to maintain the quality and real-time nature of task processing. Through this dual-focused approach, the model adeptly navigates the complex environment of resource management, striving to reduce resource consumption while meeting performance requirements.

\subsubsection{Solution Algorithm}
Given the intricacies and potential conflicts inherent in the above objective function, which is a complex combinatorial optimization problem. We recognize the dynamic of tasks and server selection variables within this problem. These dynamics make it difficult for traditional methods to address them efficiently, particularly when considering the tradeoff optimization of multiple conflicting objectives such as minimizing resource consumption while minimizing task processing delay and ensuring high accuracy. To solve these difficults, we conceptualize the optimization problem as a Combinatorial Multi-Armed Bandit (CMAB) problem \cite{chen2016combinatorial161616}. The CMAB problem reflects the multiple decision-making processes of task scheduling and resource allocation in edge-cloud systems. They are adept at handling situations characterized by uncertainty and the need for balance between exploration and exploitation. To address this CMAB problem, we propose a cross-domain collaborative optimization algorithm based on a feedback mechanism. The optimization process is shown in Algorithm 1. This algorithm is designed to iteratively improve the decision-making process by learning from the outcomes of previous allocations, thus navigating towards an approximate solution.

\begin{algorithm}
\SetAlgoLined
%\KwResult{Write here the resu }
 \textbf{Input}$:$\par
  \quad \,      The set of tasks $V$, The set of servers $Y$; \\

  \par
  
  \textbf{Output}$:$\par
 \quad \, The task and resource allocation scheme $Z^*$, $S^*$ ;

 \textbf{Procedure}$:$ \\   
    {
\ \,1: \, For each task, initialize the upload to the edge \par 
\qquad server if it is a bandwidth preference, otherwise \par \qquad initialize the upload to the cloud server.\

\ \,2: \qquad \textbf{for} each search task\

\ \,3:  \qquad \quad \,  $  t \leftarrow t+1$

\ \,4:  \qquad \quad \, Play a super arm $S_t$

\ \,5: \qquad \quad \, \textbf{if} $A_{i,t} \geq A_{i,t}^{q}$ and $D_{i,t} \leq D_{i,t}^{q}$  \textbf{then}

\ \,6: \quad \qquad \qquad   $  Z^* \leftarrow z_{i,t}^{j}=1$ \ 

\ \,7: \qquad \quad \, \textbf{else} \ 

\ \,8: \quad \qquad \qquad Update $Reg(T)$ and $S_{t}$ \ 

\ \,9: \qquad \quad \, \textbf{end if} \ 

10:\, \qquad \textbf{end for}\

11: \,  \textbf{return} $Z^*$, $S^*$;

\par
    }
 \caption{Cross-Domain Collaborative Optimization Algorithm }
\end{algorithm}

\par
The goal of the algorithm is to dynamically allocate resources to minimize resource consumption by learning and interacting with the environment. The initial step in this process involves leveraging the insights gained from the resource preference prediction module to guide the preliminary selection of servers for task allocation. Each super arm $S_t$ represents a resource allocation scheme. The super arm is instrumental in representing comprehensive resource allocation strategies that cross edge-cloud systems. The reward obtained by playing a super arm is defined as follows:

\begin{equation}
R(S)=-\frac{1}{T} \sum_{t=0}^{T} \mathbb{E}[\left(U_{t}+ \varphi  B_{t}\right)]
\end{equation}

The maximum reward is obtained by choosing different super arms. To solve the inherent challenge of attaining an exact solution to such a combinatorial optimization problem, we introduce approximation coefficients $\alpha$ and $\beta$, where  $\alpha, \beta<1$. The introduction of $\alpha$ and $\beta$ allows for the delineation of an approximate regret function. These coefficients are instrumental in guiding the algorithm toward an approximate and feasible solution by quantifying the degree of approximation acceptable in the optimization process. The approximate regret function is then defined as:
\begin{equation}
Reg(T)=T \cdot \alpha \beta \cdot R\left(S_{max}\right)-\mathbb{E}\left[\sum_{t=1}^T R\left(S_t\right)\right]
\end{equation}
where $S_{max}$ is the expected reward of the optimal super arm. The algorithm learns and finds the optimal resource allocation scheme by balancing the strategies of new super arms and the current optimal super arm. The parameters are updated through the feedback mechanism. After each selection of a super arm, it can reach at least $\alpha$ times the maximum expected reward with probability $\beta$. Finally, the resource allocation strategy is continuously adjusted through iterative training and parameter adjustment, so as to obtain the resource allocation scheme of the edge-cloud system and meet the accuracy requirement and delay requirement of the task. The feedback mechanism plays a vital role in this process, enabling the algorithm to dynamically adjust its strategy based on the observed performance of different resource allocation solutions. This adaptive approach facilitates a more effective exploration of the solution space, taking into account the changes in task requirements.

\section{Performance Evaluation}
In this section, we evaluate the performance of the CDIO framework through extensive experiments. We use real-time video inference for the object detection task to evaluate the performance of the proposed CDIO framework, and all experiments are dedicated to this task.

\subsection{Evaluation Setup}

\subsubsection{Datasets and Implementation Details} To verify the validity of CDIO, we conduct a performance evaluation on the UA-DETRAC dataset \cite{wen2020ua171717}. The UA-DETRAC dataset is a collection of video clips with a resolution of 960 × 540 collected from surveillance cameras at traffic intersections, including four weather conditions: cloudy, night, sunny, and rainy. We use four NVIDIA Jetson Xavier NX as edge devices. This edge device has a 6-core ARM v8.2 64-bit CPU and 8 GB memory. We use one NVIDIA A100 GPU with 40 GB memory as the cloud server. Referring to the experimental setup in \cite{yang2023javp666}, we simulate two network conditions with the stable bandwidth set to 300 Mbps and the fluctuating bandwidth set to change within 20\%. In addition, the above experiments are conducted under Ubuntu 18.04.3, Python 3.8.13, and PyTorch 1.11.0.

\begin{table}[th]
%\footnotesize
  \caption{The presentation of different models.}
  \label{tab1}
  \tabcolsep 12pt 
  \begin{tabular}{cccl}
    \toprule
  Model & Params (M) & mAP$_{50}$  & FLOPs  \\\hline
  YOLOv5s & 7.2  & 56.8 & \ 16.5 \\
  YOLOv5l &  46.5 & 67.3 & \ 109.1\\
  YOLOv5x &  86.7 & 68.9 & \ 205.7\\
  YOLOv5s6 &  12.6 & 63.7 & \ 16.8\\
  YOLOv5l6 &  76.8 & 71.3 & \ 111.4\\
  YOLOv5x6 &  140.7 & 72.7 & \ 209.8\\
  \bottomrule
\end{tabular}
\end{table}

\subsubsection{Models} We adopt YOLOv5 \cite{business181818} as the DNN model, which is an object detector that can be deployed on multiple types of hardware devices. We deploy small-size models on the edge device and deploy complex models on the cloud server. YOLOv5 has many different size model versions. To better validate the effectiveness of the proposed CDIO framework, we choose six models for this experiment, as shown in Table ~\ref{tab1}.  Specifically, we set up four different model deployment versions, as shown in Table ~\ref{tab2}. For version 1 and version 3, the number of model parameters on the cloud server is about 6 times that of the edge. For version 2 and version 4, the number of model parameters on the cloud server is about 12 times that of the edge.

\begin{table}[th]
%\footnotesize
  \caption{The different model deployment versions.}
  \label{tab2}
  \tabcolsep 6pt 
  \begin{tabular}{ccl}
    \toprule
  Model Deployment Versions & Edge Server & Cloud Server    \\\hline
  Model V1 & YOLOv5s  & \quad YOLOv5l  \\
  Model V2 &  YOLOv5s & \quad YOLOv5x \\
  Model V3 &  YOLOv5s6 & \quad YOLOv5l6 \\
  Model V4 &  YOLOv5s6 & \quad YOLOv5x6 \\
  \bottomrule
\end{tabular}
\end{table}

\subsubsection{Evaluation Metrics} We use end-to-end accuracy and delay as the main performance metrics. Accuracy is evaluated by the standard metric mAP$_{50}$ in object detection \cite{padilla2020survey303030},  as widely adopted in existing work. It is the mean average precision of multiple objects when the threshold of IoU is greater than 0.5. The delay includes the pre-processing delay of resource preference prediction, transmission delay, and inference delay.

\subsubsection{Baseline Methods} We compare our solution with the following three baseline methods.
\par

\begin{itemize}

\item \textbf{DAO\cite{murad2022dao999}:} This is a dynamic adaptive offloading method for video inference, which can enhance inference efficiency by dynamically adjusting network bandwidth and video bitrate.

\item \textbf{Sniper\cite{liu2022sniper888}:} This is a time-aware edge-cloud collaborative task scheduling method, which can ensure scheduling accuracy and improve throughput by the performance characterizing network. 

\item \textbf{JAVP \cite{yang2023javp666}:} This is a joint-aware video processing architecture for edge-cloud collaboration, which can guide resource allocation and reduce video processing costs by predicting the complexity of tasks. 

\end{itemize}

% \textbf{Evaluation Metrics.} We compare our solution with the following three baseline methods. We compare our solution with the following three baseline methods. We compare our solution with the following three baseline methods.

\subsection{Evaluation Results}
\subsubsection{Accuracy Analysis} We first evaluate the accuracy of CDIO in different bandwidth environments. To simulate the diversity of task processing requirements, we randomly select the accuracy requirements of the input task from [50, 80] and set the range of delay requirements to [0.2s, 0.6s], these ranges represent a wide range of applications \cite{yang2023javp666}. The experimental comparison results of accuracy under different model deployment versions are reported in Table ~\ref{tab3}. It can be found that CDIO achieves accuracy close to that of other methods. The goal of our proposed solution is not simply to improve the accuracy but to meet the task processing requirements as much as possible. Therefore, CDIO suffers some loss in average accuracy, which is in line with our expectations.

Next, we measure the success rate in meeting the predefined accuracy requirements for tasks. Specifically, a task is adjudged successful if its final processing accuracy surpasses the input accuracy requirements. Conversely, a task failing to meet these criteria is categorized as unsuccessful.  The test results show that CDIO has a success rate of 93\% at stable bandwidths. Remarkably, even under fluctuating bandwidth conditions, CDIO maintains a high success rate of 92\%. This performance is significantly superior to the success rates of other methods, which are observed to only range between 80\% and 87\%. CDIO achieves the highest success rate under different bandwidth conditions and has significant advantages. The advantage is more obvious under fluctuating bandwidths. This shows that it can adapt to different task requirements through resource preference prediction.

\begin{table}[th]
  \caption{The accuracy comparison results under different methods.}
  \label{tab3}
  \begin{tabular}{c|ccccl}
    \toprule
    \multicolumn{2}{c}{Method}&DAO &Sniper& JAVP &CDIO\\
    \midrule
    \multirow{3}*{ } 
            &Model V1 & 62.23 & 62.81 & 62.65 & 62.79\\
    Stable    &Model V2 & 63.15 & 63.68 & 63.43 & 63.57\\
       Bandwidths       &Model V3 & 67.18 & 67.73 & 67.46 & 67.63\\
                &Model V4 & 67.74 & 68.51 & 68.27 & 68.44\\
       \midrule
    \multirow{3}*{ } 
             &Model V1  & 61.64& 62.18 & 62.43 & 62.31\\
    Fluctuating  &Model V2 & 62.13 & 62.86  & 63.21 & 63.15\\
   Bandwidths &Model V3 & 66.79& 67.29& 67.37 & 67.35\\
   &Model V4 & 67.21 & 67.87 & 68.17 & 68.12\\
  \bottomrule
\end{tabular}
\end{table}

\subsubsection{Delay Analysis} 
\begin{figure}[th]
	\centering 
    \subfigure[Stable Bandwidths]{
    \label{Fig.sub.5.1}
    \includegraphics[width=0.22\textwidth]{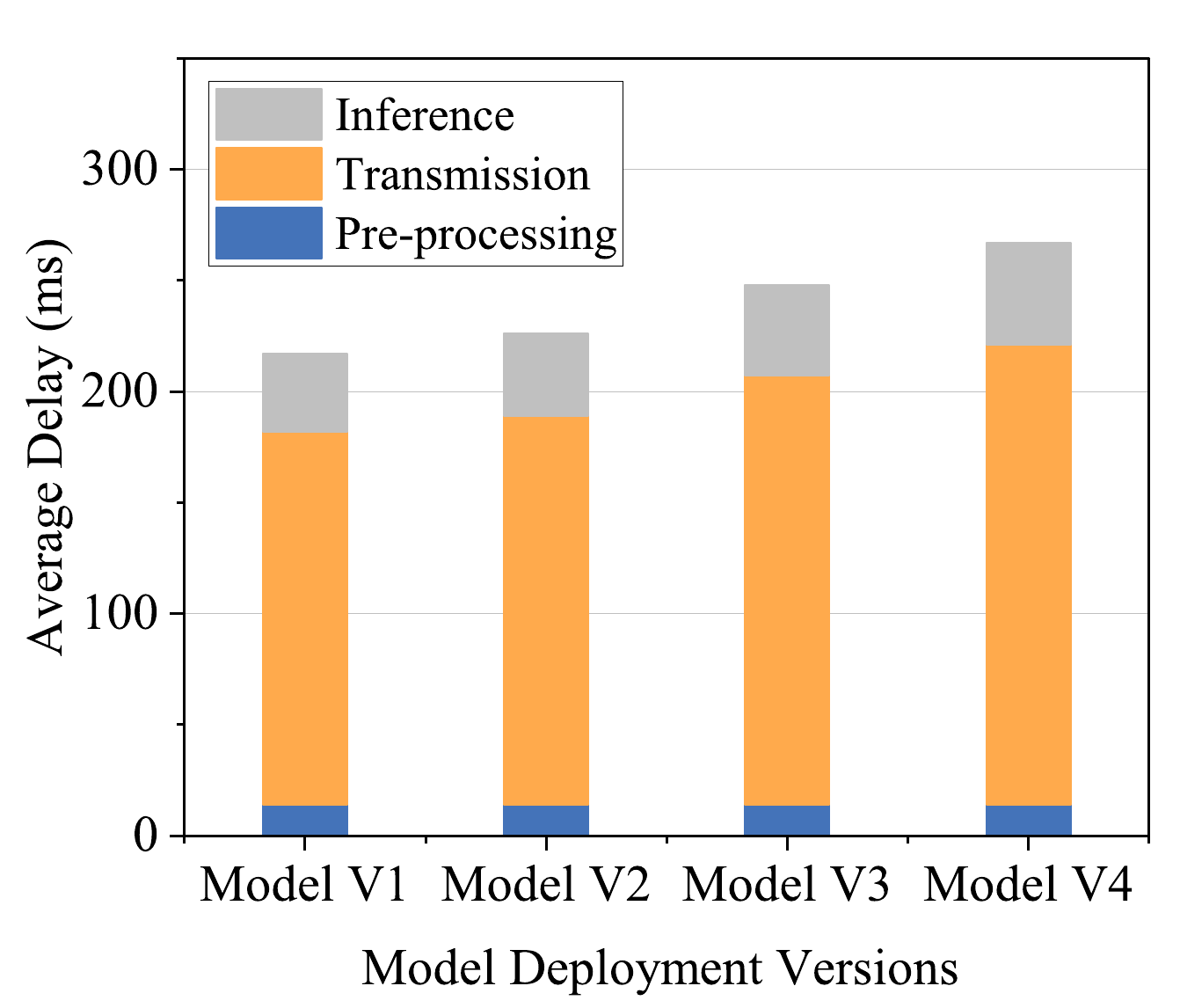}}
    \subfigure[Fluctuating Bandwidths]{
    \label{Fig.sub.5.2}
    \includegraphics[width=0.22\textwidth]{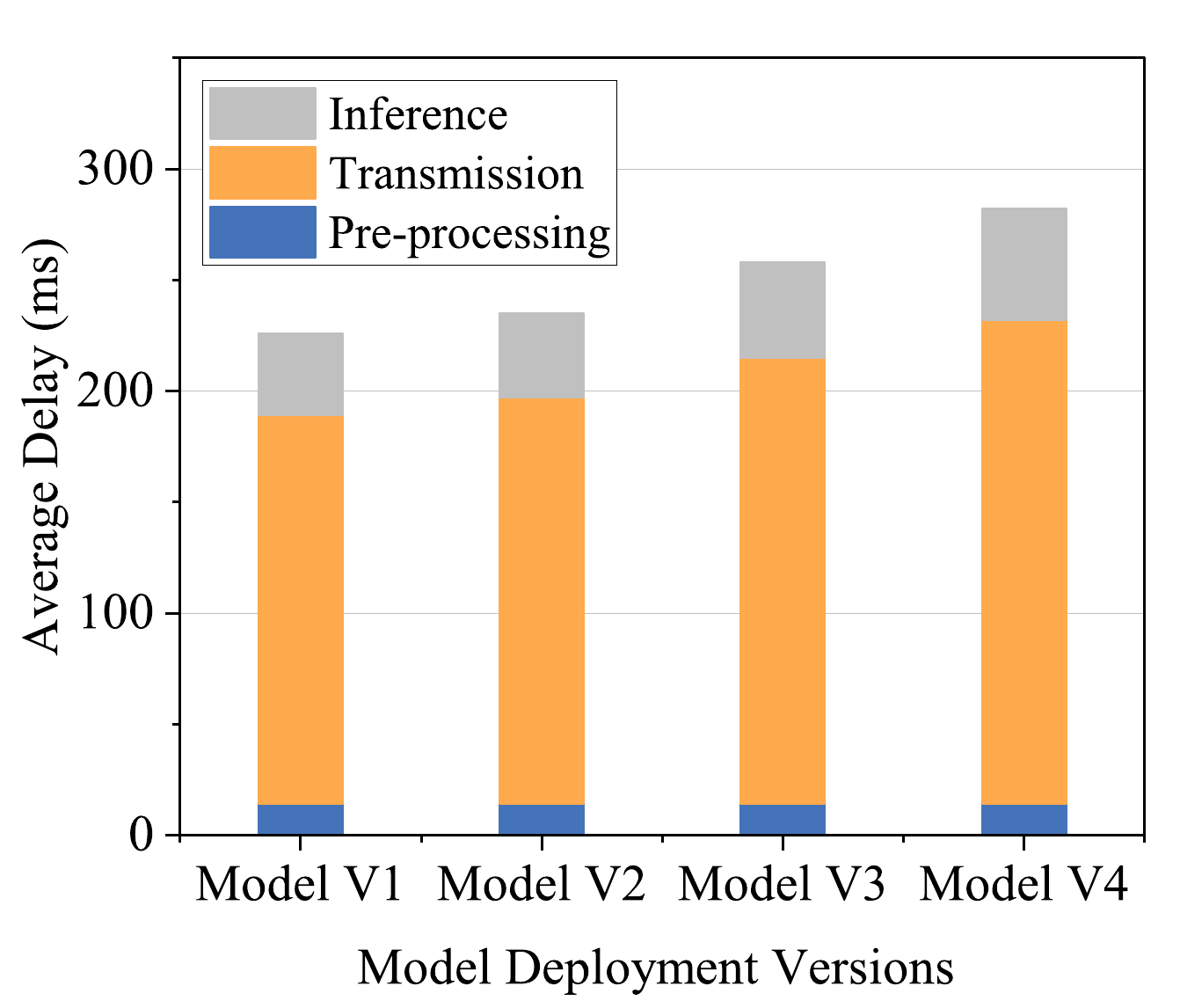}}
    \caption{The delay analysis within CDIO framework.}
	\label{figure5}
\end{figure}

The delay of the resource preference prediction module is crucial to the overall system performance. We aim for it to be able to distinguish different types of tasks in real-time and avoid resource waste. Therefore, we first test the running time of the resource preference prediction module. The results show that the pro-processing delay of resource preference prediction is 14ms.  Figure~\ref{figure5} illustrates the average delay of CDIO, including pre-processing delay, transmission delay, and inference delay. We can find that the delay in predicting resource preferences for tasks is very low, which has little impact on the overall system delay. For delay requirements, CDIO achieves a success rate of 90\%-94\%, and other methods are only 74\%-86\%. It provides a solid foundation for cross-domain resource optimization in edge-cloud systems.

\par
\begin{figure}[th]
	\centering 
    \subfigure[Stable Bandwidths]{
    \label{Fig.sub.6.1}
    \includegraphics[width=0.22\textwidth]{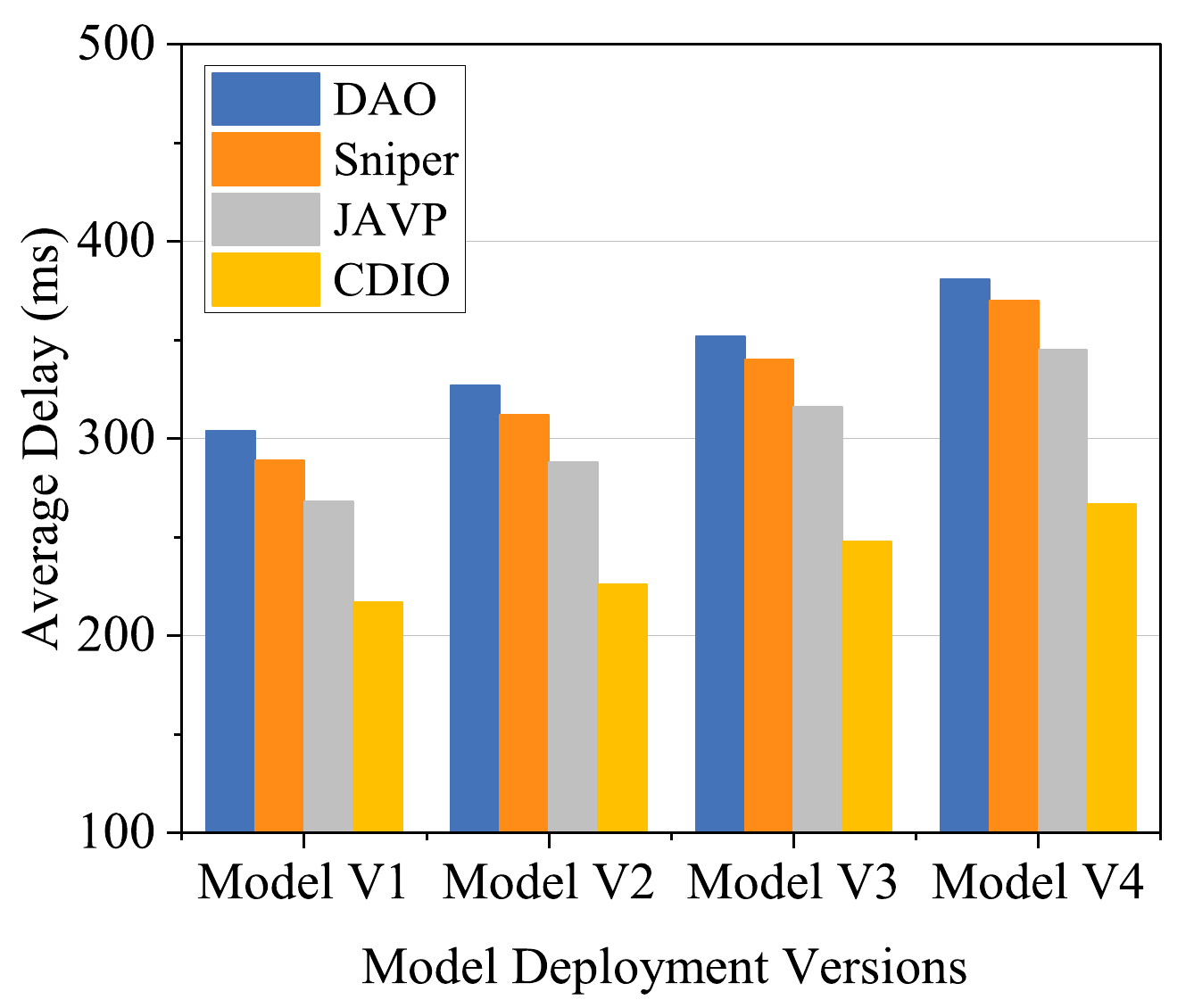}}
    \subfigure[Fluctuating Bandwidths]{
    \label{Fig.sub.6.2}
    \includegraphics[width=0.22\textwidth]{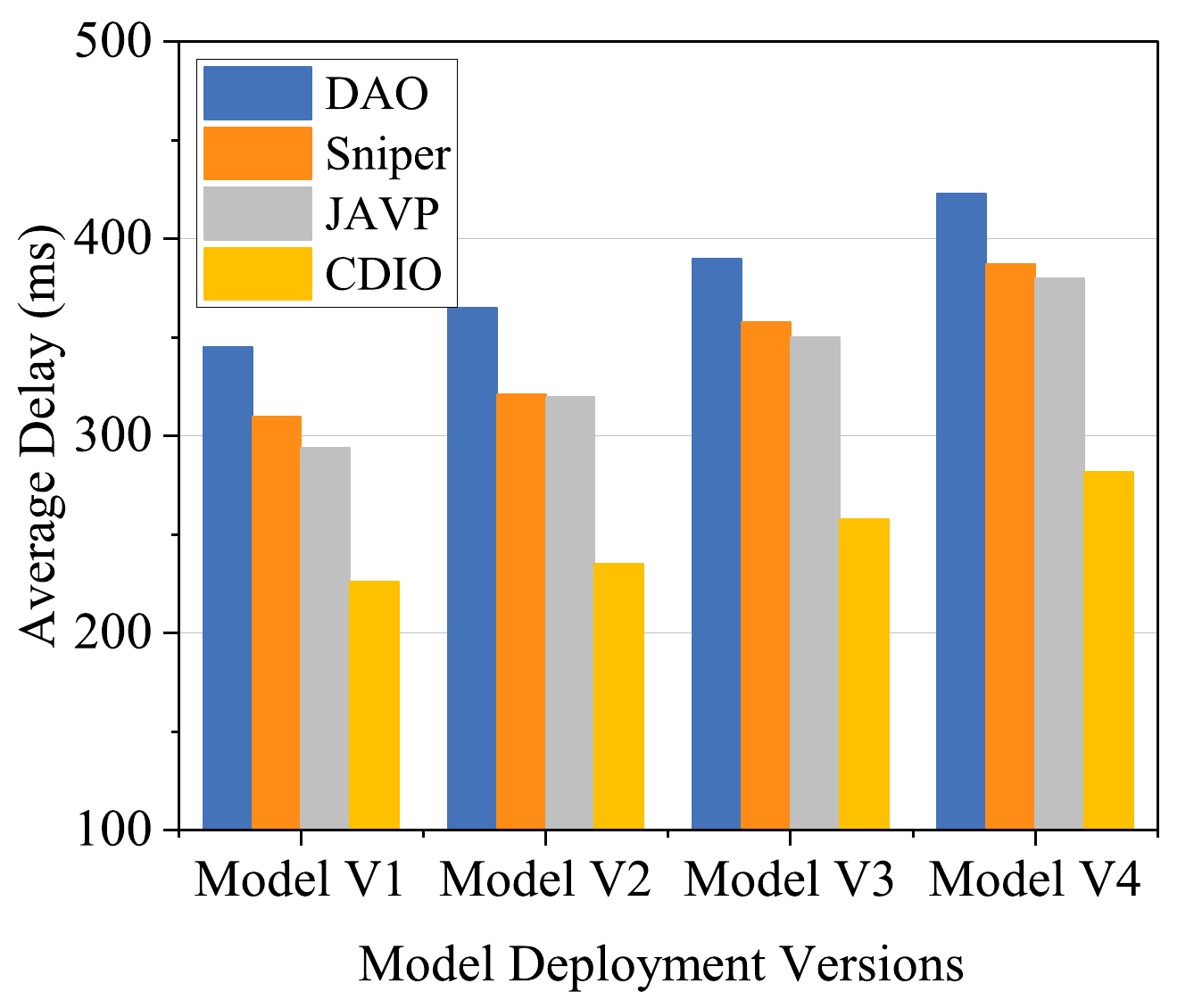}}
    \caption{The comparison results of average delay under different methods.}
	\label{figure6}
\end{figure}

Figure~\ref{figure6} shows the comparison results for delay under different methods, where CDIO achieves the lowest delay.  It can be found that the delay of CDIO is lower than the other methods for different model deployment versions. The advantage is more obvious in fluctuating bandwidth environments. The reason why CDIO can achieve lower delay and maintain high accuracy is that the dynamic allocation of resources can have a significant impact on the delay for task processing. The cross-domain collaborative optimization module within CDIO can adeptly modulate resource distribution in response to delay requirements and improve accuracy.

\subsubsection{Computing Consumption Comparison} We further evaluate the computing consumption of CDIO under different bandwidth environments. The computing consumption is mainly evaluated by the product of the server's half-precision computational performance (FP16) and the average inference time required for processing tasks. The FP16 computational performance is a measure of how many trillion floating-point operations per second (T Flops) the server can handle. Inference time is obtained by testing the running time of the task inference process. By multiplying the FP16 performance by the average inference time, we arrive at a nuanced understanding of the computing consumption for different methods.

\begin{figure}[th]
	\centering 
    \subfigure[Stable Bandwidths]{
    \label{Fig.sub.7.1}
    \includegraphics[width=0.22\textwidth]{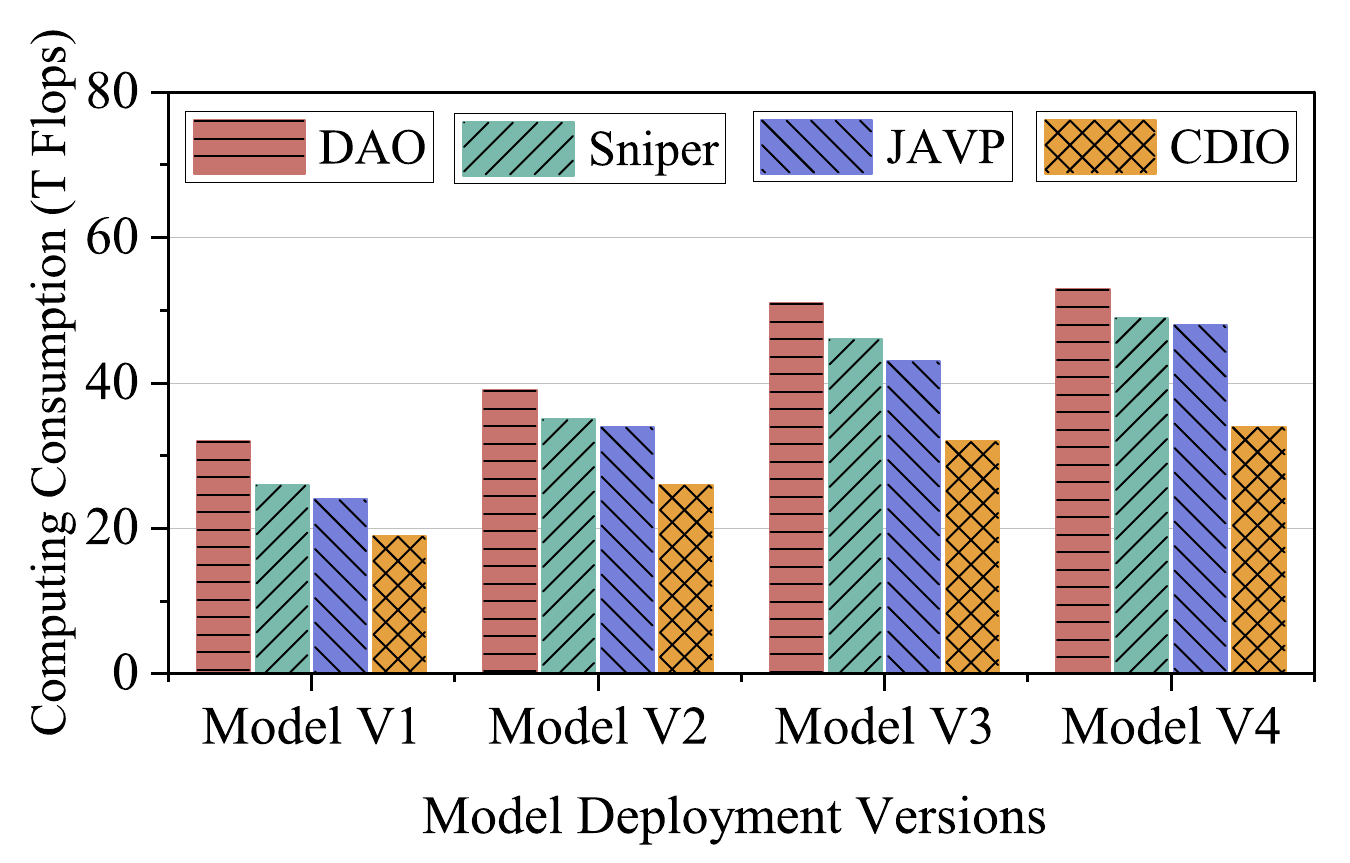}}
    \subfigure[Fluctuating Bandwidths]{
    \label{Fig.sub.7.2}
    \includegraphics[width=0.22\textwidth]{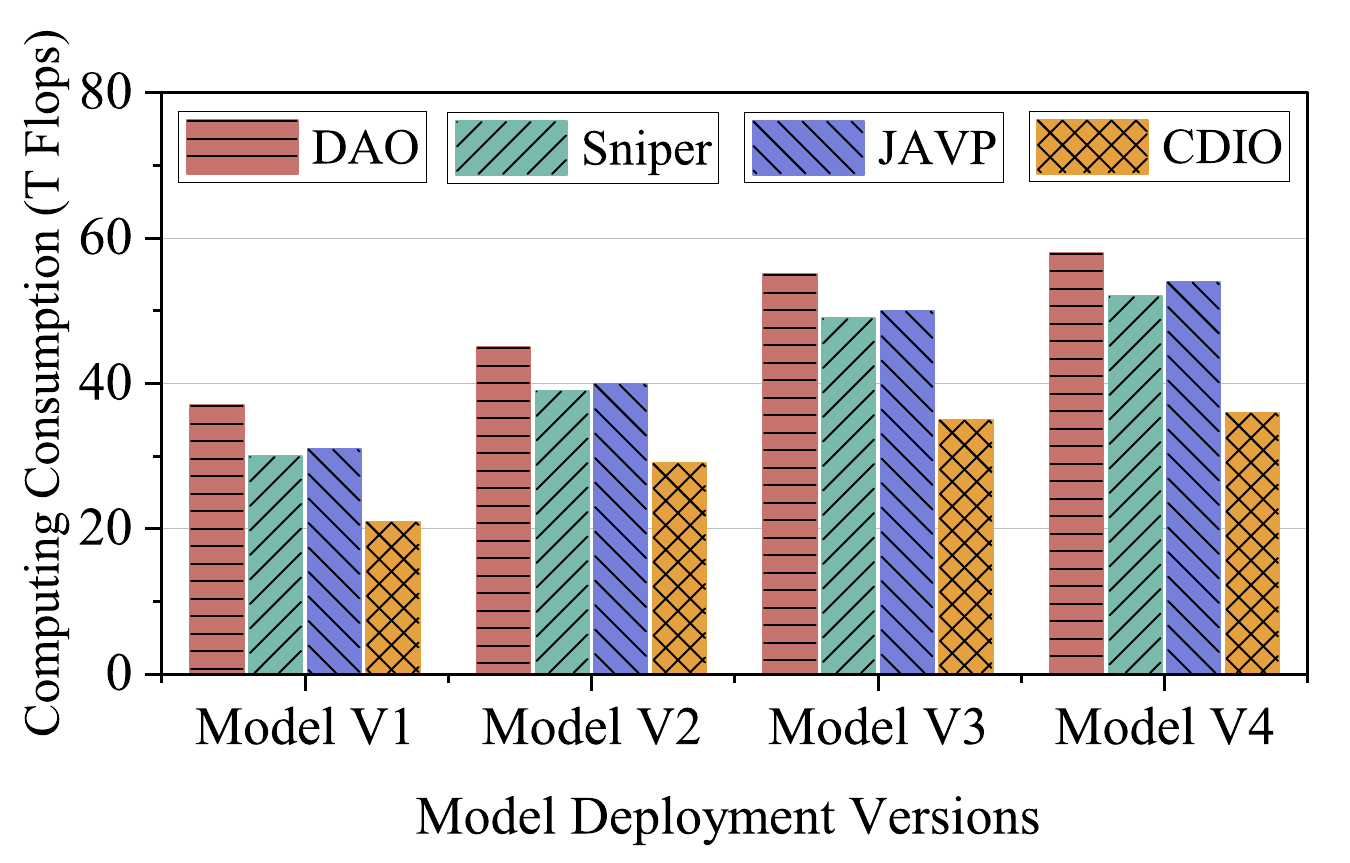}}
    \caption{The comparison results of computing consumption under different methods.}
	\label{figure7}
\end{figure}

\par
Figure~\ref{figure7} illustrates the comparison results for computing consumption. We find that CDIO achieves the best performance under different model deployment versions. CDIO also achieves better results under fluctuating bandwidths. This is because resource preference prediction and cross-domain collaborative optimization are more easily adapted to dynamic environments. It enables CDIO to adjust more effectively to changes in resource availability and task requirements. Specifically, CDIO exhibits an average reduction in computing consumption of over 20\% when compared to the JAVP and Sniper, and an even greater reduction of over 30\% relative to the DAO. This can indicate that CDIO has a better capacity to deliver high performance while minimizing computing consumption.

\subsubsection{Bandwidth Consumption Comparison} The bandwidth consumption is a crucial measure for assessing network efficiency and resource usage. We further evaluate the bandwidth consumption of the proposed CDIO framework. The bandwidth consumption is mainly evaluated by dividing the amount of data transmitted by the task by the transmission time. To determine this time, we use precise timing mechanisms that begin timing at the initiation of the data transfer and stop once the last packet of data has been successfully received. By dividing the total amount of data transmitted by the transmission time, we obtain a concrete value for the bandwidth consumption. 

\par
\begin{figure}[th]
	\centering 
    \subfigure[Stable Bandwidths]{
    \label{Fig.sub.8.1}
    \includegraphics[width=0.22\textwidth]{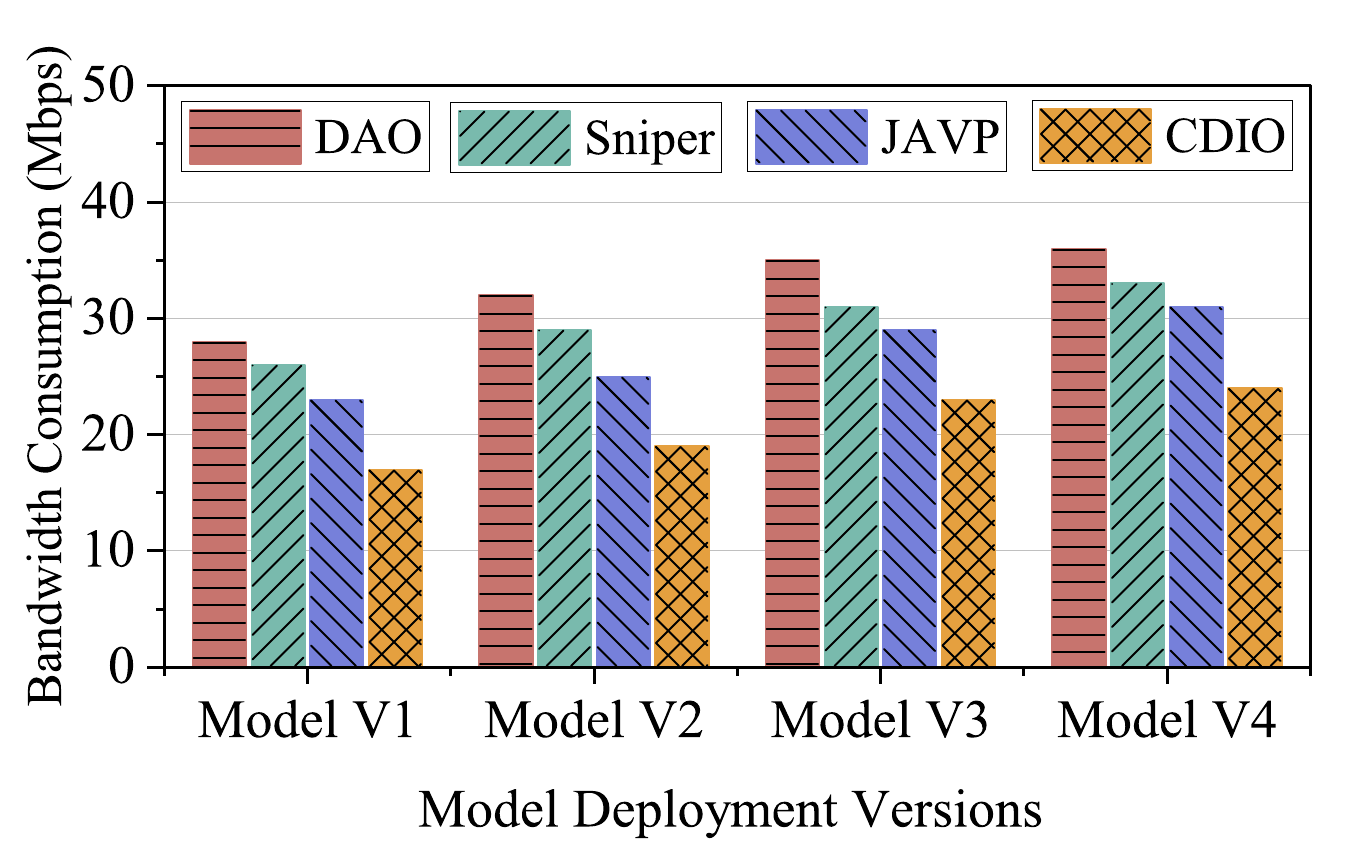}}
    \subfigure[Fluctuating Bandwidths]{
    \label{Fig.sub.8.2}
    \includegraphics[width=0.22\textwidth]{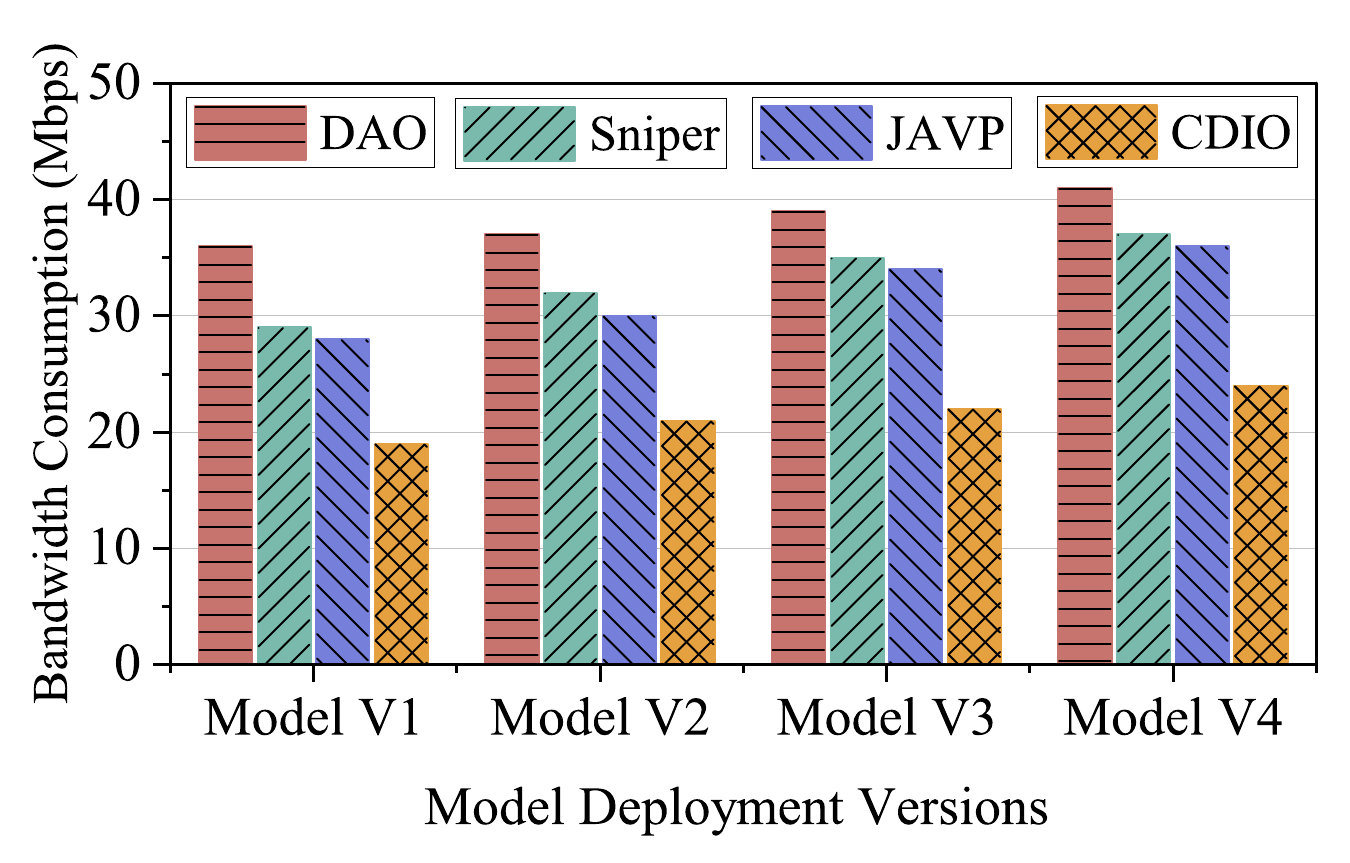}}
    \caption{The comparison results of bandwidth consumption under different methods.}
	\label{figure8}
\end{figure}

Figure~\ref{figure8} illustrates the comparison results for bandwidth consumption under different methods. By looking at Figure~\ref{Fig.sub.8.1} and Figure~\ref{Fig.sub.8.2}, we find that CDIO significantly outperforms other methods in terms of bandwidth consumption both in stable and fluctuating bandwidths. On average, CDIO reduces bandwidth consumption by 20\%-40\% compared to other methods. This reduction shows that CDIO can minimize unnecessary data transmission and optimize network resource allocation. Whether in conditions of stable bandwidths or fluctuating bandwidths, CDIO can ensure robust data transmission efficiency. This can also enhance the overall performance of systems deployed within bandwidth-constrained environments.

\subsubsection{Energy Consumption Comparison} 
\begin{figure}[th]
	\centering 
    \subfigure[Stable Bandwidths]{
    \label{Fig.sub.9.1}
    \includegraphics[width=0.22\textwidth]{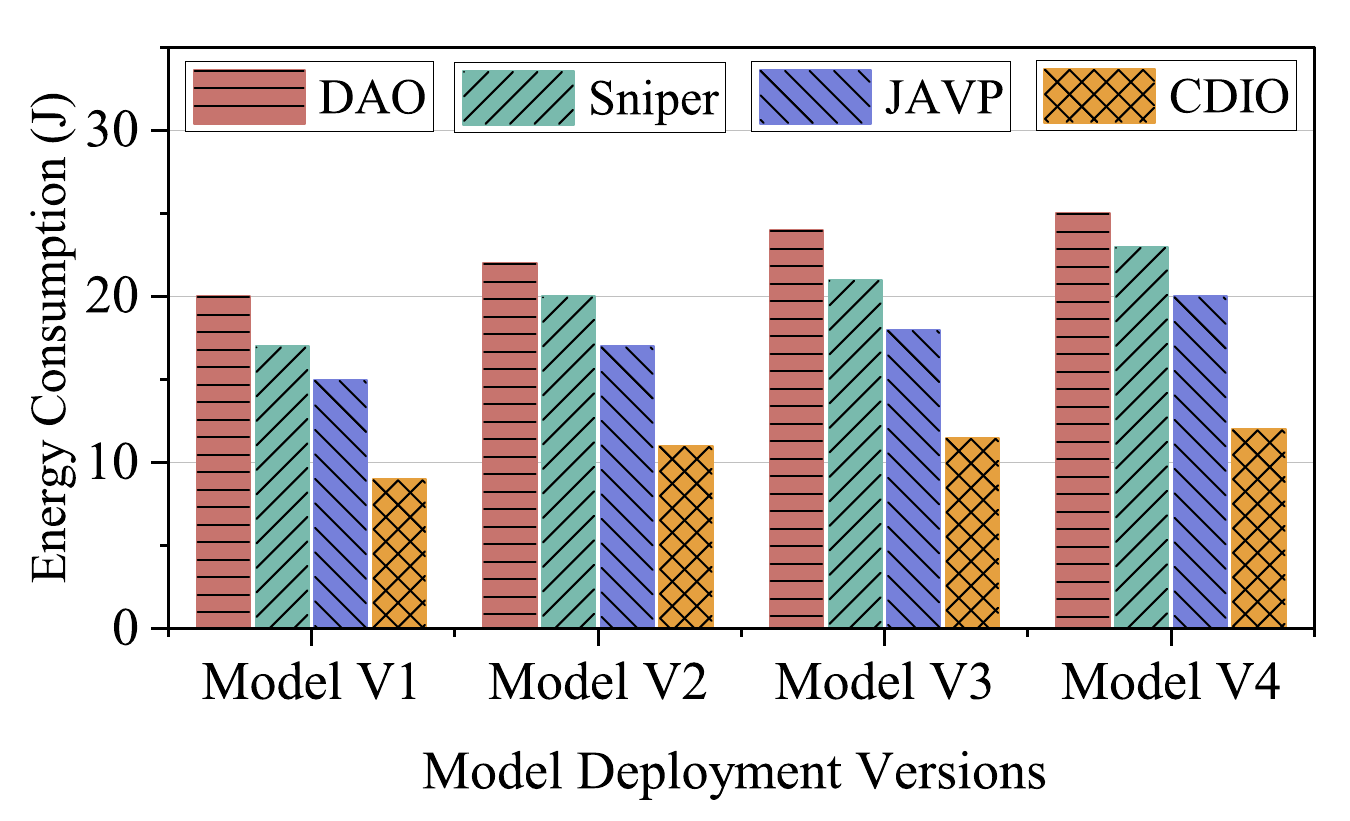}}
    \subfigure[Fluctuating Bandwidths]{
    \label{Fig.sub.9.2}
    \includegraphics[width=0.22\textwidth]{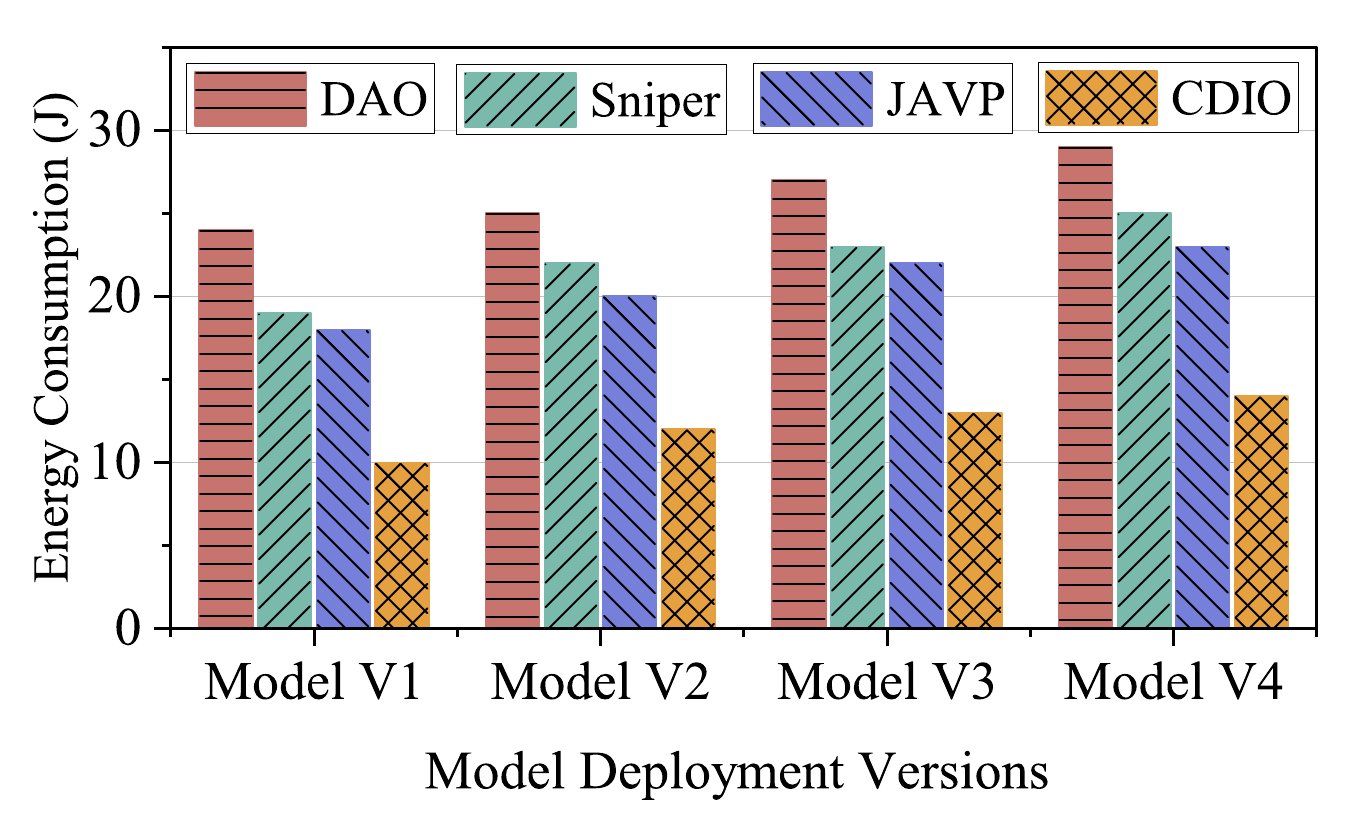}}
    \caption{The comparison results of energy consumption under different methods.}
	\label{figure9}
\end{figure}

Finally, we evaluate the energy consumption of the different methods. In our setup, the energy consumption includes server working energy consumption, server idle energy consumption, and transmission energy. The working energy consumption is calculated by multiplying the server's working power by the processing time of the task, and the 
idle energy consumption is calculated by multiplying the server's idle power by the idle time. The transmission energy consumption is obtained by multiplying transmission power by the transmission time.
Figure~\ref{figure9} illustrates the comparison results for energy consumption under different methods. CDIO adapts well to dynamic scenes, consistently demonstrating lower energy consumption than other methods. On average, CDIO can reduce energy consumption by more than 40\%. This is because CDIO can rationally schedule tasks and improve resource utilization through resource preference prediction and cross-domain collaborative optimization. Overall, CDIO mitigates the environmental impact associated with high energy consumption in edge-cloud systems. This cross-domain optimization ensures that CDIO can be adaptable to a wide array of application scenarios in the future.

\subsubsection{Ablation Studies} To deeply analyze the solution proposed in this paper, we present the results of ablation experiments. These experiments are designed to evaluate the individual and combined contributions of the core components of our proposed framework. The ablation studies include three different experimental schemes: (1) only the resource preference prediction (RPP) module, (2) only the cross-domain collaborative optimization (CDCO) module, and (3) the RPP module + CDCO module. The comparison metrics include accuracy, delay, computing consumption, bandwidth consumption, and energy consumption. 
\begin{table}[th]
  \caption{The test results with different modules of the CDIO framework.}
  \label{tab4}
  \begin{tabular}{ccccl}
    \toprule
    Method &RPP &CDCO &RPP+CDCO\\
    \midrule
     Accuracy & 65.13 & 61.37 & 65.42 \\
    Delay & 297 & 314 & 245\\
    Computing Consumption & 33.5 & 30.8 & 29.1\\
   Bandwidth Consumption & 24.6 & 22.4 & 21.2\\
    Energy Consumption & 14.1 & 13.9 & 11.6\\
  \bottomrule
\end{tabular}
\end{table}

\par
The RPP module's primary function is to predict the most suitable resources for given tasks, aiming to optimize the allocation process based on predicted preferences. The CDCO module is designed to coordinate resources and tasks across different devices, facilitating a more efficient utilization of available resources. By conducting ablation experiments on different modules of the proposed method, we find that the RPP module plays a key role in improving the accuracy. Table~\ref{tab4} shows the average results. If only the CDCO module is used,  the accuracy decreases by about 4\%. If only the RPP module is used, the computing consumption and bandwidth consumption grow by about 15\%. This suggests that CDCO is critical in reducing resource consumption. In addition, both the RPP module and the CDCO module play a central role in reducing delay and energy consumption. If we only adopt the RPP module or CDCO module, the delay and energy consumption increase significantly. This dual contribution is indicative of the collaborative relationship between the RPP module and the CDCO module, each complementing the other to enhance the system’s overall efficiency. In conclusion, the findings from these experiments are quite revealing. This indicates that both the RPP and CDCO modules are integral to the effectiveness and generalization capability of our proposed method.

\section{Discussion}
\textbf{Impact on Multimedia Systems.} CDIO is designed to enhance the efficiency of edge-cloud collaboration in handling vast and diverse video tasks. By predicting resource preference types through an analysis of spatial complexity and processing requirements, CDIO allows for a more intelligent allocation of resources. This is critical in multimedia systems, where the diversity of tasks demands flexible and efficient processing capabilities. It then employs a cross-domain collaborative optimization algorithm that intelligently guides the allocation of resources within the edge-cloud system. This approach ensures that resources are utilized more efficiently, paving the way for enhanced performance and sustainability in future multimedia systems.  As multimedia systems continue to evolve, CDIO's resource preference prediction module and cross-domain collaborative optimization module ensure that the framework can adapt to new technologies and requirements. It potentially sets a new trend for how video tasks are managed and executed in edge-cloud systems.

\par
\textbf{Advantages Analysis.} CDIO is an edge-cloud collaborative inference framework that supports resource preference prediction. Video tasks come with a wide range of requirements, from simple to highly complex. By predicting the specific resource preference types of each task, the system can ensure that each task is matched with the most appropriate resources. This targeted allocation significantly reduces the problem of over-provisioning or under-utilizing resources. The availability and condition of resources in edge-cloud systems can vary widely over time. Predicting resource preference types allows the system to dynamically adjust to these fluctuations, ensuring that tasks are always allocated to the most efficient resources currently available. This adaptability is crucial for maintaining high levels of system efficiency despite the inherent variability in resource availability.

\par
\textbf{Limitations Analysis.} CDIO in the current version has some limitations. Firstly, despite its impressive success rates of more than 90\% under stable bandwidths and fluctuating bandwidths, there remains a notable proportion of tasks that fail to meet the accuracy and delay requirements. The reasons behind these failures could vary from inaccuracies in predicting resource preferences, to limitations in resource allocation algorithms, or unforeseen complexities in the tasks themselves. Secondly, the CDIO framework does not yet support memory optimization. In the future, we will work to address the above limitations.

\section{Conclusion}
In this paper, we present CDIO, the edge-cloud collaborative inference framework that supports resource preference prediction. It can dynamically decide task allocation strategies and guide resource optimization according to the type of resource preference of tasks and current system resource usage. The main goal of CDIO is to realize adaptive scheduling for diverse tasks and improve inference efficiency under dynamic resource conditions. Compared with state-of-the-art solutions, experimental results show that CDIO can significantly reduce resource consumption and improve the performance of video task processing in edge-cloud systems. Future work will incorporate memory and focus on multi-dimensional resource collaborative optimization in edge-cloud systems.


\bibliographystyle{ACM-Reference-Format}
\bibliography{sample-base}

\appendix



% Lorem ipsum dolor sit amet, 


% Etiam commodo feugiat nisl pulvinar pellentesque. Etiam


% pulvinar ex pretium. Sed interdum orci quis metus 

\end{document}